\documentclass[twocolumn]{aa}
\usepackage{graphicx}
\usepackage{psfig}
\usepackage{graphicx}
\begin{document}

\title{Infrared spectroscopy of a brown dwarf companion candidate
near the young star GSC~08047-00232 in Horologium\thanks{Based on
observations obtained on Cerro Paranal, Chile, in ESO programs 65.L-0144
and 71.C-0206C} }

\author{R. Neuh\"auser \inst{1} \and E.W. Guenther \inst{2}}

\offprints{Ralph Neuh\"auser, rne@astro.uni-jena.de}

\institute{Astrophysikalisches Institut, Universit\"at Jena, 
Schillerg\"a{\ss}chen 2-3, D-07745 Jena, Germany \\
\and Th\"uringer Landessternwarte Tautenburg, Sternwarte 5, D-07778 Tautenburg, Germany }

\date{received 19 Nov 2003; accepted 3 March 2004}

\abstract{We present infrared H- and K-band spectra of 
a companion candidate $3^{\prime \prime}$ north of the young star GSC~08047-00232, 
a probable member of the nearby young Horologium association.
From previously obtained JHK-band colors and the magnitude difference between primary and 
companion candidate, the latter could well be substellar (Neuh\"auser et al. 2003)
with the spectral type being roughly M7-L9 from the JHK colors (Chauvin et al. 2003).
With the H- and K-band spectra now obtained with ISAAC at the VLT, 
the spectral type of the companion candidate is found to be M6-9.5.
Assuming the same age and distance as for the primary star ($\sim 35$ Myrs, 50 to 85 pc),
this yields a mass of $\sim 25$ Jupiter masses for the companion,
hence indeed substellar. After TWA-5 B and HR 7329 B, this is the third
brown dwarf companion around a nearby ($\le 100$~pc) young ($\le 100$~Myrs) star.
A total of three confirmed brown dwarf companions (any mass, 
separation $\ge 50$ AU)
around 79 stars surveyed in three young nearby associations 
corresponds to a frequency of
$6 \pm 4~\%$
(with a correction for missing companions which
are almost on the same line-of-sight as the primary star
instead of being separated well),
consistent with the expectation, if binaries have the same 
mass function as field stars. Hence, it seems that
there is no brown dwarf desert at wide separations.
}

\maketitle

\markboth{Neuh\"auser \& Guenther: Brown dwarf companion to GSC~08047-00232}{}

\section{Introduction: Young brown dwarf companions}

Brown dwarfs (BDs) are objects with mass below the hydrogen burning mass limit,
i.e. below $\sim 0.078$~M$_{\odot}$ (e.g. Burrows et al. 1997).
It is not yet clear, how BDs form.
According to one of the suggested scenarios, accreting companions 
get ejected as BD embryos by encounters inside a forming multiple system 
(e.g. Reipurth \& Clarke 2001).
In such a case, BDs may be different from stars regarding 
multiplicity, kinematics, and disk properties,
so that such a scenario can be tested observationaly,
%
%
e.g. by comparing the secular evolution of 
star-star binaries with star-BD binaries: If single BDs became single by the
ejection from a multiple system, then the fraction of star-BD binaries should
decrease faster with time than star-star binaries.

A possible secular evolution in stellar binaries has been studied by
Bouvier et al. (2001) and Patience et al. (2002) with as yet inconclusive
results. In order to compare such results with star-BD binary fractions,
one would first need significant statistics.
Hence, several groups search for BDs as companions to stars.

As far as young stars ($\le 100$ Myrs) in nearby associations ($\le 100$ pc) 
are concerned, only two brown dwarfs have been confirmed so far as companions 
by both common proper motion as well as spectroscopy, 
namely TWA-5~B (Lowrance et al. 1999, Neuh\"auser et al. 2000) 
in the TW Hya association (TWA, Webb et al. 1999)
and HR~7329~B (Lowrance et al. 2000, Guenther et al. 2001) 
in the Horologium association (HorA, Torres et al. 2000) 
and/or $\beta$ Pic moving group (Zuckerman \& Webb 2000).

In Sect. 2, we present previous and new imaging observation
and investigate the astrometry, i.e. whether the pair is co-moving.
Our spectroscopic observations, the data reduction,
and the final H- and K-band spectra are shown in Sect. 3. 
We obtain a mass estimate for the companion and
interprete the results in Sect. 4.
A discussion in terms of BD companion frequency
is given in the last section.

\section{Imaging observation: A companion candidate}

Recently, both Neuh\"auser et al. (2003, henceforth N03) and 
Chauvin et al. (2003, henceforth C03) have searched for faint
companions to confirmed and probable
members of the young Horologium and Tucana associations,
which may form one common association (TucHorA). N03 used the normal
infrared (IR) imaging camera SOFI and the MPE speckle camera Sharp,
both at the ESO-3.5m-NTT, while C03 used the ADONIS AO system at the ESO-3.6m 
telescope. Their samples are partly overlaping and partly disjunct, 
e.g. N03 also included $\beta$ Pic members.
Both groups detected a promising companion candidate $3^{\prime \prime}$
north of GSC~08047-00232, a K3~V star suggested to be a Horologium member
by Torres et al. (2000) with H$\alpha$ emission 
and a Lithium equivalent width of 0.35~\AA , 
located at $\alpha = 01^{h} 52^{m} 14.63^{s}$
and $\delta = -52^{\circ} 19^{\prime} 33.1^{\prime \prime}$ (J2000.0);
the star is also called TYC 8047-232-1 (Tycho, Hog et al. 2000), 
ERX 14 (Torres et al. 2000),
and 1RXSJ015215.7-521939 (ROSAT, Voges et al. 1999).
The position in the sky, the radial velocity, the proper motion,
and the kinematical parallaxe (and its position in the H-R diagram
at that distance) are all consistent with membership to HorA 
(Torres et al. 2000) and also with the possible common TucHor association.
Also, the youth indicators like H$\alpha$ emission, Lithium absorption,
X-ray emission, and rotational velocity are also consistent qualitativelly 
and quantitativelly with an age of roughly 35 Myrs as HorA/TucHorA (Torres et al. 2000).

Given the JHK-band colors, C03 found that the companion candidate could
have a spectral type between M7 and L9 by comparing the Leggett et al. (2002)
photometry with the Geballe et al. (2002) spectral classifications.
From the magnitude difference in K, also N03 concluded that the
companion candidate could be a BD companion of late-M or L spectral type.

\begin{figure}
\vbox{\psfig{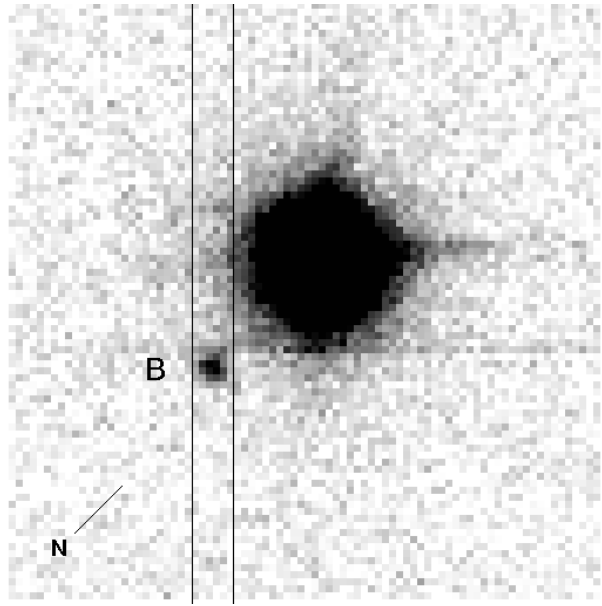}}
\caption{Aquisition image through a narrow band filter in the K-band
taken with ISAAC at the VLT before our spectroscopy,
showing the companion candidate B three arc sec north of GSC 08047-00232 A.
The $0.8^{\prime \prime}$ wide slit was located neither along both objects 
nor perpendicular to that direction, but inbetween the two, namely from
SE (top) to NW (bottom) with the faint companion candidate in the slit and the 
bright primary star outside the slit, so that we still obtain some stray light 
from the bright star in the slit, which we use for calibration purposes.
North is to the bottom left, the line is 1.5 arc sec long.}
\end{figure}

To confirm or reject its companion status, 
one can either try to (i) take a spectrum,
or (ii) prove the pair to be co-moving,
or (iii) show the radial velocity or astrometric
wobble of the primary due to the companion (both very small),
or (iv) detect orbital motion (after a long time).
Only the first option is immediately possible to achieve,
given the fact that the proper motion of the primary star is very
small ($47.0 \pm 2.4$ mas/yr), almost exactly to the east,
so that the separation to the companion candidate ($3^{\prime \prime}$ north)
would hardly change, even if it is a background object.

\begin{table*}
\begin{tabular}{llcccccl}
\multicolumn{8}{c}{\bf Table 1. Astrometry and photometry} \\ \hline
Date    & Instrument & Sep. $[\prime \prime]$ & PA [deg]          & J [mag]          & H [mag]         & K [mag] & Ref. \\ \hline
04 July 2001 & NTT-Sharp   & $3.238 \pm 0.022$ & $357.65 \pm 0.18$ &                  &                 & $15.0 \pm 0.3$ & N03 \\
29 Oct 2001 & 3.6m-Adonis & $3.210 \pm 0.118$ & $359.2 \pm 2.3$   & $16.25 \pm 0.25$ & $15.2 \pm 0.18$ & $14.9 \pm 0.2$ & C03 \\
20 July 2003 & VLT-ISAAC  & $3.138 \pm 0.048$ & $\sim 358$ & \multicolumn{3}{l}{(narrow band filter)} & here \\ \hline
\end{tabular}
\end{table*}

Before obtaining a spectrum, we took aquisition images,
which can be used to measure the separation between primary
star and companion candidate. This separation should be
constant with time, if its a common proper motion pair;
a change in separation due to orbital motion in case
of a non-circular orbit would be negligible.

The companion candidate is detected marginally in the 
aquisition image on 10 July 2003, but detected 
clearly on 20 July 2003 (Fig. 1), obtained through
a narrow band filter at $2.19~\mu m$ in the K-band.

We compile the previous and new astrometric results in Table 1.
There are images at three epochs available now, first detected
on 04 July 2001 (N03), then on 29 Oct 2001 (C03), and finally
on 20 July 2003 (here).
We measure the position of the primary star and the companion
candidate in every co-added image with center/gauss in MIDAS, 
which provides $1~\sigma$ errors.
Pixel scales, detector orientation, and their errors are measured on 
calibration fields with many stars of known fixed positions taken 
in the same nights.\footnote{While the measurement of the separation
between the two object is very precise on the ISAAC aquisition
image obtained in July 2003, we could not measure the position angle PA
with high precision, see Table 1, because there are no other objects
detected on the (short exposure) aquisition image, with whose
positions we could have measured the actual detector orientation;
hence, we have to rely on the header information, which we know
to be less precise from previous experience; also, being just
an aquisition image obtained in a service mode night, we do not
have other images obtained at the same detector orientation
of other fields, which could otherwise be used for such a calibration;
hence, the large uncertainty in the third PA values in Table 1,
while the separation is more precise being independant of the
detector orientation.}
With these values, we can obtain separations and position
angles (PA), see Table 1.

Our three measurements -- separated by two years - of the separation 
and the PA are consistent with eachother within $1.5~\sigma$.
The companion candidate is located 
almost directly north of the primary star 
(only 1 to $2^{\circ}$ west of the north direction)
and the primary star is known to move towards the east (just a bit south of east).
Hence, the separation should increase with time, if the companion
is a non-moving background star. Given the proper motion of the 
primary star, it should have increased to $3.25 ^{\prime \prime}$
by July 2003. As observed (Table 1), the separation 
has not changed within the errors.
The astrometry is $2~\sigma$ deviant from a background hypothesis.
However, due to the small proper motion and the small epoch difference,
we cannot yet reliably distinguish between the companion candidate being
either a common proper motion companion and a background star.

\section{Spectroscopic observations}

We obtained H- and K-band spectra with the Infrared Spectrometer
And Array Camera (ISAAC)
at Antu, Unit Telescope 1 (UT1) of the Very Large Telescope (VLT) of the
European Southern Observatory (ESO) on Cerro Paranal, Chile.
ISAAC is equipped with a $1024 \times 1024$ HAWAII detector.
All H- and K-band spectra obtained have 60 sec exposure each and are taken 
through a $0.8^{\prime \prime}$ slit with a low resolution of $R \simeq 1700$.
In the H-band, centered on $1.65~\mu m$, we took 28 spectra on 20 July 2003.
In the K-band, centered on $2.20~\mu m$, we took 28 spectra on 11 July 2003
and again 28 spectra on 20 July 2003. All data were obtained in service mode.
Calibration frames were taken in the same nights: darks, flats, and
arcs for wavelength calibration.

We measured the full width at half maximum (FWHM) on the aquisition images
as well as in the individual spectra obtaining $0.5^{\prime \prime}$ for 
both nights. The S/N in the (unbinned) K-band spectrum taken on 20 July 2003 
is $\sim 24$, but only $\sim 20$ in the data taken on 11 July 2003,
so that we use the K-band spectrum from 20 July 2003.

\begin{figure}
\vbox{\psfig{figure=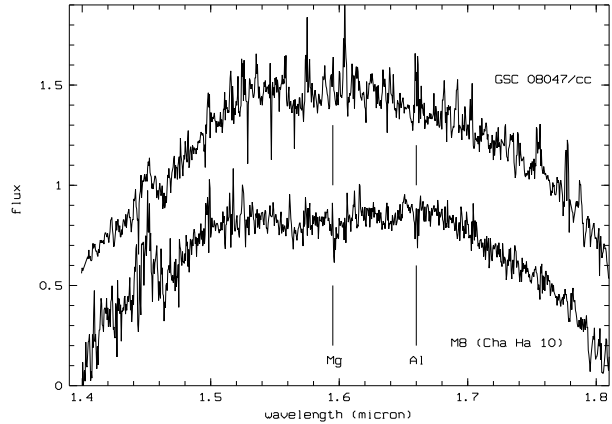,width=9cm,height=6cm,angle=270}}
\caption{H-band spectra of the companion candidate GSC 08047-00232/cc (top) 
compared to the H-band spectrum of the M7.5-type young BDs Cha H$\alpha$ 10
(bottom) in Chamaeleon also observed by us with ISAAC with the same set-up
(in June 2000). 
Both spectra are relatively featureless and show a similar
continuum shape, so that the companion is 
similar to Cha H$\alpha$ 10 in its spectral type 
($\pm 1$ sub-class, i.e. M6.5-8.5).}
\end{figure}

\begin{figure}
\vbox{\psfig{figure=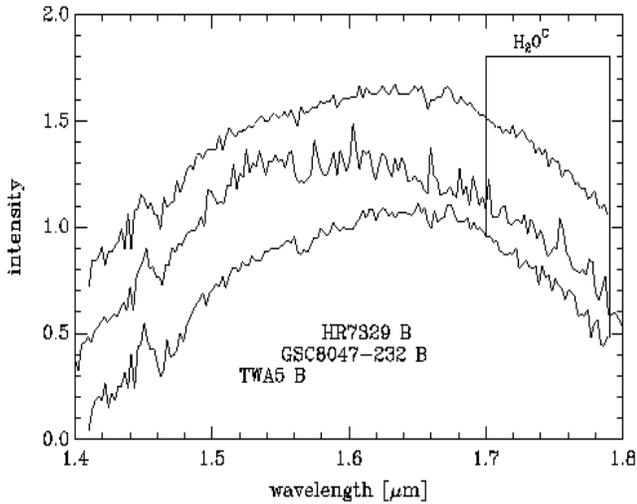,width=9cm,height=7cm,angle=0}}
\caption{H-band spectra of GSC 08047-00232/cc (middle) compared
with HR 7329 B (M7-8, top) and TWA-5 B (M8.5-9, bottom).
Here, we have smoothed the spectral resolution by a blocking factor of 5
yielding a resolution of $\sim 350$ and a S/N ratio of $\sim 40$.
The slope in the red part of the spectrum of GSC 08047-00232/cc
is intermediate between that of HR 7329 B and TWA-5 B.
The Reid et al. (2001) spectral index H$_{2}$O-C is most sensitive
to the continuum slope and, hence, spectral type, indicated in the upper right,
yielding roughly M8 as spectral type for GSC 08047-00232/cc.}
\end{figure}

Data reduction was done in the normal way with IRAF:
Dark subtraction, normalization, flat fielding, sky subtraction,
wavelength calibration, co-adding the spectra, then
correction for instrumental sensitivity and atmospheric response.
We use the spectrum of the bright primary star 
(located off the slit) for calibration purposes. 
The S/N in the binned spectra is $\sim 40$ for both H and K.
For the new companion candidate as well as for the spectral templates 
HR 7329 and TWA 5 with their known companions, both the primary and the 
secondary are observed simultaneously, so that we can do almost perfect 
calibration of the continuum shapes by using the known spectral type
of the primary for estimating the response of the detector and, hence,
calibrating the secondary. This is different for the
spectral template Cha H$\alpha$ 10.
However, even with templates observed simultaneously,
some uncertainties in the calibration remain,
so that spectral types obtained from spectral indices should 
be more accurate than those obtained from continuum slopes.

\begin{figure}
\vbox{\psfig{figure=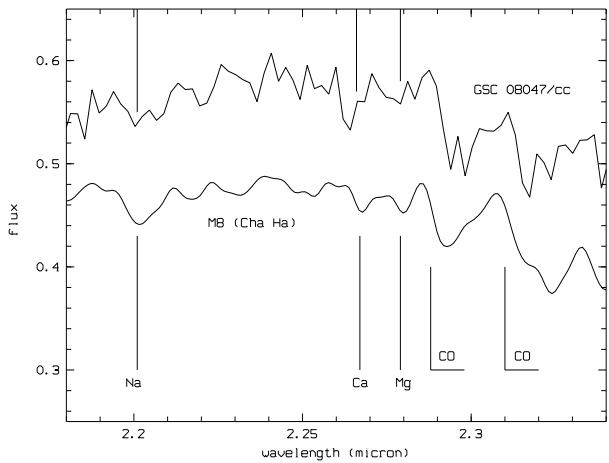,width=9cm,height=6.5cm,angle=270}}
\caption{Rebinned K-band spectra of the companion candidate GSC 08047-00232/cc (top)
compared to the averaged K-band spectrum of M7.5- to M8-type
young BDs (bottom) in Chamaeleon (from Comer\'on et al. 2000),
obtained with SOFI, an IR spectrograph similar to ISAAC,
but at a different spectral resolution.
Both spectra show the same absorption lines Na, Ca, and Mg,
very similar molecular CO bands, and the same
continuum shape, so that the companion is 
similar in spectral type as the template, i.e. M7-8.5.}
\end{figure}

The final co-added H- and K-band spectra are shown in Fig. 2 to 4,
respectively. We compare the spectra of GSC 08047-00232/cc with
those of young late-M type BDs in Chamaeleon, Tucana, and TWA.

The H-band spectrum is compared with Cha H$\alpha$ 10 
observed with ISAAC by us with the same set-up (in June 2000).
Cha H$\alpha$ 10 is a young ($\sim 2$ Myrs) M7.5 BD in 
the Cha I star forming region (Comer\'on et al. 2000).
Both spectra are relatively featureless with the same
continuum shape indicative of a very similar spectral type.
The slope in the red part of the H-band at $\sim 1.7$ to $1.8~\mu m$
is very sensitive to the temperature at late M and L spectral
types due to water absorption. As seen in Fig. 2, the red slope 
in GSC 08047-00232/cc is slightly steeper than in Cha H$\alpha$ 10;
hence, GSC 08047-00232/cc appears slightly cooler, i.e. slightly later,
around M8-8.5,
while the broad maximum of the flux in GSC 08047-00232/cc is at a 
slightly bluer wavelength compared to Cha H$\alpha$ 10, 
which would indicate a slightly earlier spectral type,
around M6.5-7.
Hence, we classify the companion candidate tentativelly as 
M6.5-8.5 (from the H-band continuum slopes).

Then, we estimate the spectral index H$_{2}$O-B following Reid et al. (2001),
the ratio of the fluxes at $1.48$ and $1.60~\mu m$.
We obtain $0.88 \pm 0.09$ for GSC 08047-00232/cc, hence M8.5-9.5.
Then, the spectral index H$_{2}$O-C is sensitive for the slope
in the red part of the H-band from 1.7 to 1.8 $\mu m$, 
i.e. very sensitive to the spectral type; 
we obtain $0.56$ for GSC 08047-00232/cc, 
$0.55$ for TWA-5~B (M8.5-9), 
and $0.58$ for HR 7329 B (M7-8).
Hence, GSC 08047-00232/cc is intermediate between TWA-5~B and
HR 7329 B and can be classified as M8-8.5 (from H$_{2}$O-C).
Together with the above comparison of GSC 08047-00232/cc and 
the M7.5-type Cha H$\alpha$ 10 yielding 
M6.5-8.5 (Fig. 2), 
we classify GSC 08047-00232/cc as M$8 \pm 2$, 
from the H-band.

The K-band spectrum (Fig. 4) is compared with an averaged K-band spectrum of 
M7.5- to M8-type BDs in Chamaeleon (Comer\'on et al. 1999, 2000) obtained with 
SOFI, because we did not take a K-band spectrum of Cha H$\alpha$ 10 or any 
other M6-8 type Chamaeleon BD with ISAAC.
Here, we see again a very similar continuum shape as well
as several lines (Na, Ca, and Mg) similar in both spectra.
Also, the CO molecular absorption bands are very similar
in both spectra, so that we can classify GSC 08047-00232/cc
as roughly M8 from the K-band, consistent with the H-band.
The K-band spectrum of GSC 08047-00232/cc is clearly later than
the average M6 and average M8-type spectra from Comer\'on et al. (1999).

To summarize, the companion candidate can be classified as 
M$8 \pm 2$ (or M6-M9.5)
from the H- and K-band spectra.
This classification is tentative and rough, because of the
relatively low spectral resolution, low S/N ratio.

\section{Result: A brown dwarf companion}

Obviously, it is very unlikely to find by chance an 
object with spectral type M$8 \pm 2$
just $3^{\prime \prime}$ off one of our targets. 
According to the Besancon galactic model, to find an object as faint as
GSC 08047-00232/cc within $3^{\prime \prime}$ of the primary star at
the given galactic latitude is only $\sim 0.4~\%$ (C03) with sample
sizes being 25 stars in N03 and 23 stars in C03 with 13 overlaps; it is even
less likely, to find such an object with (at least roughly) the same proper
motion (see Table 1) and the late M spectral type.
Hence, the companion candidate most certainly is a truely bound companion.
For the remainder of the paper, we regard this companion candidate
as a bound companion of the star and call it GSC 08047-00232~B.

The primary star GSC 08047-00232~A, as a probable
member of $\sim 35$ Myrs young TucHorA,
has a distance of $\sim 50$ pc (Torres et al. 2000, Zuckerman \& Webb 2000) 
to as much as $\sim 85$ pc (C03). The larger distance is obtained,
when forcing the star to be co-eval with the rest of the TucHorA stars
at $\sim 35$ Myrs with the Baraffe et al. (1998) models using the
hydrostatic scale-height $1.0~H_{p}$ as mixing length parameter. 
However, with e.g. $1.9~H_{p}$, the star is again older. 
On D'Antona \& Mazzitelli (1998) tracks, the star is 
30 to 50 Myrs young at 85 pc, see Fig. 5.
Hence, we assume 85 pc as distance 
and $\sim 35$ Myrs as age for both the primary star and its companion.

\begin{figure*}
\vbox{\psfig{figure=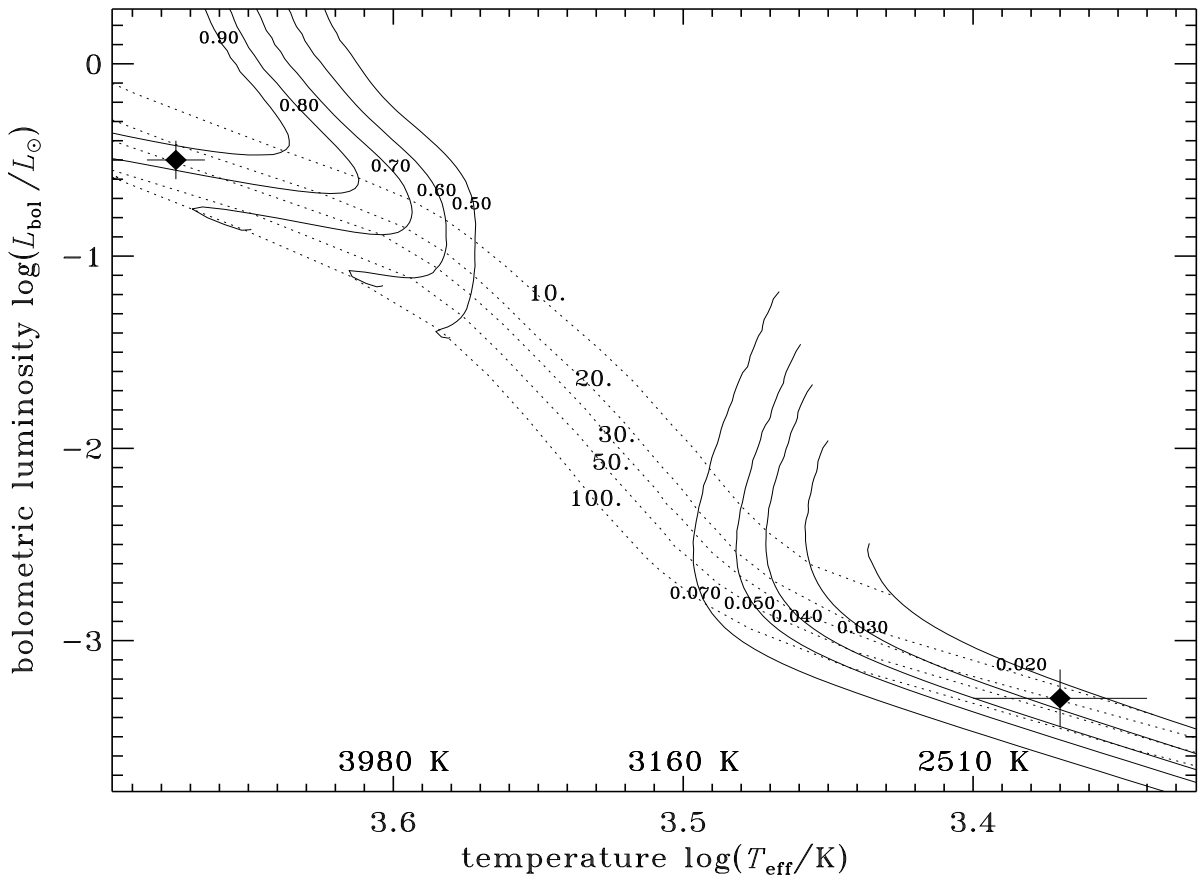,width=18cm,height=13cm,angle=90}}
\caption{H-R diagram for GSC 08047-0232 A and B (at 85 pc) with theoretical 
tracks and isochrones from D'Antona \& Mazzitelli (1998), ages in Myrs, 
masses in solar masses. The objects are co-eval at 30 to 50 Myrs.}
\end{figure*}

Given the spectral type M$8 \pm2$
(Sect. 3), we obtain $\sim 2100$ K
as temperature following Jones \& Tsuji (1997) and Kirkpatrick et al. (1999),
or $\sim 2500$ K from the Luhman (1999) extrapolation of
the Leggett et al. (1996) scale for dwarfs,\footnote{Because the
object is close to the main sequence, it is much closer to true
dwarfs than to objects intermediate between dwarfs and giants,
so that we use the dwarf scale instead of Luhman's intermediate scale;
for intermediate objects, the temperature would be $\sim 2635$ K.} 
so that we use $2300 \pm 250$ K as temperature.

The companion has an absolute magnitude of M$_{K} = 10.2 \pm 0.1$ at 85 pc.
According to $J-K=1.35 \pm 0.32$ mag (C03) and the spectral type M6-9.5,
we use a bolometric correction in the K-band of $BC_{K} = 3.0 \pm 0.1$ mag 
following Leggett et al. (2002). With $K = 14.9 \pm 0.1$ mag (C03, N03),
we then obtain a bolometric luminosity of 
$\log~(L_{bol}/L_{\odot})=-3.3 \pm 0.1$ for 85 pc
(or $-3.85 \pm 0.10$ for 50 pc).
With these values for temperature, luminosity, and age, we can obtain a 
mass estimate using theoretical model calculations, see Table 2.

Using figures 7 in Burrows et al. (1997) for $\log (L_{bol}/L_{\odot}) = -3.3$
at 35 Myrs, we obtain $\sim 20$ to 30~M$_{jup}$,
from figure 9 for $2300 \pm 250$ K at 35 Myrs, we derive $\sim 15$ to 35~M$_{jup}$,
and from figure 11 for $\log (L_{bol}L_{\odot}) = -3.3$ and $2300 \pm 250$ K,
we get $\sim 20$ to 40~M$_{jup}$.
Using figure 2 (35 Myrs) 
in Chabrier et al. (2000) with T=$2300 \pm 250$ K at 35 Myrs,
we obtain $\sim 20$ to 40~M$_{jup}$;
from table 1 in Chabrier et al. (2000) (i.e. for 100 Myrs as their youngest age) 
and using T=$2300 \pm 250$ K, $\log (L_{bol}/L_{\odot}) = -3.3$, and
M$_{K}=10.2 \pm 0.1$ mag, we obtain $\sim 35$ to 45~M$_{jup}$.
From figure 2 in Baraffe et al. (2002), 
using L and T as above at 35 Myrs, 
we obtain a mass of $\sim 20$~M$_{jup}$.
On D'Antona \& Mazzitelli (1998) tracks and isochrones, 
the primary star and the companion are co-eval at 
85 pc with an age of $\sim 20$ to 50 Myrs and
with a mass of $\sim 20$ to 50~M$_{jup}$ for the companion 
(see Fig. 5). 

\begin{table}
\begin{tabular}{lll}
\multicolumn{3}{c}{\bf Table 2. Mass determination for the companion} \\ \hline
Model                  & parameters used        & mass \\ \hline
Burr. et al. 97 fig. 7 & 35 Myrs, L at 85 pc    & 20-30 M$_{j}$ \\
Burr. et al. 97 fig. 9 & T$=2300$~K, 35 Myrs    & 15-35 M$_{j}$ \\
Burr. et al. 97 fig. 11& T$=2300$~K, L at 85 pc & 20-40 M$_{j}$ \\
Chab. et al. 00 fig. 2 & T$=2300$~K, 35 Myrs    & 20-40 M$_{j}$ \\
Chab. et al. 00 tab. 1 & T, L \& M$_{K}$ at 85 pc & 35-45 M$_{j}$ \\
Bara. et al. 02 fig. 2 & T, 35 Myrs, L at 85 pc & $\sim 20$~M$_{j}$ \\
Bara./Chab. in C03     & M$_{K}$ vs. J-K CMD    & 20-40 M$_{j}$ \\
Wuch. \& Tsch. 2003    & T, L at 85 pc, 35 Myrs & $\sim 7$~M$_{j}$ \\
D'Ant. \& Mazz. 98     & T, L at 85 pc (fig. 5) & 20-50 $M_{j}$ \\ \hline
\end{tabular}
\end{table}

C03 show a color-magnitude diagram (CMD), M$_{K}$ versus $J-K$ (their figure 4),
for the primary star with its companion candidate; they compare their locations 
in the CMD at 60 and 85 pc distances with model calculations made by 
Baraffe et al. (1998) and Chabrier et al. (2000) for 20 to 50 Myrs;
they find the two objects to be co-eval for 85 pc only (at $\sim 40$ Myrs)
with the mass of the companion then being $\sim 20$ to 40~M$_{jup}$.
Hence, with both the D'Antona \& Mazzitelli (1998) model and the
Baraffe et al. (1998) model, we consistently obtain $\sim 35$ Myrs
and $\sim 30$~M$_{jup}$ for the companion.

We have to caution, however, as discussed in Wuchterl (2001) and Baraffe et al. 
(2002), that theoretical models are more uncertain for young ages (even tens of Myrs)
as compared to older (Gyr) ages, because of uncertain initial conditions.
Hence, there could easily be a systematic shift in our mass estimates.
Extrapolating the Wuchterl \& Tscharnuter (2003) models by scaling the luminosity 
dependance in mass and age (to 35 Myrs), one would obtain a mass estimate of 
only $\sim 7$~M$_{jup}$, below the 
deuterium burning mass limit (G. Wuchterl, private communication),
which we will study and discuss in more detail elsewhere.

To summarize, we obtain a mass estimate of 
$\sim 15$ to 50~M$_{jup}$ with $\sim 25$~M$_{jup}$
as most likely value, from conventional models by the 
D'Antona \& Mazzitelli, Chabrier \& Baraffe, as well as Burrows et al. groups,
but possibly a lower mass when formation scenarios are taken into account.
GSC~08047-00232~B is probably a low-mass brown dwarf.
To confirm companionship of this late-M type object with the primary star beyond any 
doubt, common proper (or radial) motion (or reflex motion of the primary due to the 
companion) or orbital motion ($\sim 3000$ year orbit with $\sim 250$ AU semi-major axis) 
would have to be shown with high significance.

If the primary star GSC 08047-00232 would not be a member of HorA or TucHorA,
the assumed distance for primary and companion could be wrong. However,
the strong Lithium absorption line (together with the other youth indicators
like strong X-ray emission, H$\alpha$ emission, and fast rotation) clearly
indicate that the primary star is a pre-main sequence star, 
so that the spectral type of the companion, 
M$8 \pm2$, alone points to a mass near or below the H burning limit.

\section{The frequency of brown dwarf companions}

Most young stars in the nearby TWA, TucHorA, and $\beta$ Pic associations 
have been surveyed by deep IR imaging (with speckle and/or AO and/or HST)
for substellar companions, so that we can try to do some statistics:

In TWA, there are 19 member systems (see Torres et al. 2003 for a recent list)
including visual multiples, so that there is a total number of 27 stars.
All of them have been surveyed by now with one BD companion found and
confirmed, namely TWA-5~B (Lowrance et al. 1999, Neuh\"auser et al. 2000).
The sensitivity of the surveys are deep enough to have detected all
BD companions down to the D burning limit outside of roughly 50 AU;
however, they are quite inhomogenous, done by several groups with
several different telescopes and instruments.
There is a total of one BD companion among 27 stars,
which is a frequency of $4 \pm 4~\%$.

With imaging, one can detect only those companions separated from the
primary in the two dimensions of the plane of the sky,
i.e. those separated well in right ascension $\alpha$ and/or declination $\delta$.
One cannot detect those companions directly (or almost directly)
in front of or behind the primary, i.e. located (almost) on the same 
line-of-sight as the primary, instead of having an angular separation.
By direct imaging, one can detect companions only in the two dimensions 
$\alpha$ and $\delta$, not in the 3rd dimension.\footnote{In radial
velocity searches, one can detect all companions, unless the orbital
plane is perpendicular to the line-of-sight.}
To correct the number above (from two to three dimenions, 2D to 3D),
we have to add $\sim 1/2$ of the 2D value; hence, we obtain $\sim 6 \pm 6~\%$
(for TWA).

In the $\beta$ Pic association, Neuh\"auser et al. (2003) have done a
homogenous survey for companions down to the D burning limit outside
of $\sim 50$ AU. Before, one BD companion was known, namely HR~7329~B
(Lowrance et al. 2000, Guenther et al. 2001). Neuh\"auser et al. (2003)
did not find any additional BD companion (or candidates) among
a total of 17 stars in 12 systems.
Hence, the frequency of BD companions in $\beta$ Pic is $6 \pm 6~\%$
(or $9 \pm 9~\%$ after correcting from 2D to 3D).

In TucHorA, 25 stars were observed by Neuh\"auser et al. (2003) and
ten more by Chauvin et al. (2003), both down to roughly the D burning limit 
for companions outside of $\sim 50$ AU. The number of BD companions here
is again one, namely the one confirmed above.
Hence, here we have a frequency of BD companions of $3 \pm 3~\%$
(or $4 \pm 4~\%$ for 3D).

In all three associations, the frequency of BD companions (of any mass
down to the D burning limit of $\sim 13$~M$_{jup}$) outside of roughly
50 AU separation is very low, around a few per cent.
The spectral types of the primaries ranges from A0 (HR~7329~A)
over K3 (here, GSC~08047-00232~A) to M2 (TWA-5~A).
Its still low number statistics with one BD companion each,
but since 17 to 35 stars have been observed, one can clearly 
say that the frequency of wide BD companions ($\ge 50$ AU) is quite low.
Putting together the three associations, we have three BDs around
79 stars, i.e. a frequency of $4 \pm 3~\%$ 
(or $6 \pm 4~\%$ after correction of the projection effect from 2D to 3D), 
which is the 
best estimate so far for the frequency of BDs (of any mass) outside of 50 AU 
around $\sim 10$ to $35$ Myrs young stars born in loose associations.
This value is consistent with the expectation, if the mass function of binaries
is similar to the mass function of field stars, in which case $\sim 3$ to $12~\%$ of
stars should have BD companions (see Gizis et al. 2001), 
so that there seems to be
no evidence for a BD desert at wide separation.

\acknowledgements{We would like to thank the ESO Service Mode observing team
on Cerro Paranal as well as the ESO User Support Group for perfect support
in both observing periods 65 and 71.
M. Ammler and T. Sch\"oning prepared figure 5.
We also acknowledge Wolfgang Brandner and G\"unther Wuchterl for many stimulating 
discussions.}

{}


\begin{thebibliography}{}
\bibitem[2002]{baraf}
Baraffe I., Chabrier G., Allard F., Hauschildt P.H., 2002, A\&A 382, 563
\bibitem[2001]{bouv01}
Bouvier J., Duchene G., Mermilliod J.C., Simon T., 2001, A\&A 375, 989
\bibitem[1997]{burr97}
Burrows A., Marley M., Hubbard W.B., et al., 1997, ApJ 491, 856
\bibitem[2000]{chab2000}
Chabrier G., Baraffe I., Allard F., Hauschildt P.H., 2000, ApJ 542, 464
\bibitem[2003]{chauv03}
Chauvin G., Thomson M., Dumas C., et al., 2003, A\&A 404, 157 (C03)
\bibitem[1999]{com1999}
Comer\'on F., Rieke G.H., Neuh\"auser R., 1999, A\&A 343, 477
\bibitem[2000]{com00}
Comer\'on F., Neuh\"auser R., Kaas A.A., 2000, A\&A 359, 269
\bibitem[1998]{dam98}
D'Antona F., Mazzitelli I., 1998, {\em A Role for Superadiabatic Convection in Low Mass Structures~?} 
In: {\it Brown dwarfs and extrasolar planets} Rebolo R., Martin E.L., Zapatero-Osorio M.R. (Eds.),
Proceedings of a Workshop held in Puerto de la Cruz, Tenerife, Spain, 17-21 March 1997, 
ASP Conf. Ser. 134, p. 442
\bibitem[2002]{geb02}
Geballe T.R., Knapp G.R., Leggett S.K., et al., 2002, ApJ 564, 466
\bibitem[2001]{gizis}
Gizis J.E., Kirkpatrick J.D., Burgasser A., Reid I.N., Monet D.G., Liebert J., Wilson J.C.,
2001 ApJ 551, L163
\bibitem[2001]{guenther}
Guenther E.W., Neuh\"auser R., Hu\'elamo N., Brandner W., Alves J., 2001, A\&A 365, 514
\bibitem[2000]{hog}
Hog E., Fabricius C., Makarov V.V., et al., A\&A 355, L27 
\bibitem[1997]{jotsu97}
Jones H.R.A., Tsuji T., 1997, ApJ 480, L39
\bibitem[1995]{kenyon}
Kenyon S.J., Hartmann L.W., 1995, ApJS 101, 117
\bibitem[1999]{kirk99}
Kirkpatrick J.D., Reid I.N., Liebert J., et al., 1999, ApJ 519, 802
\bibitem[1996]{leg96}
Leggett S.K., Allard F., Berriman G., Dahn C.C., Hauschildt P.H., 1996, ApJS 104, 117
\bibitem[2002]{leg02}
Leggett S.K., Golimowski D.A., Fan X., et al., 2002, ApJ 564, 452
\bibitem[2000]{low}
Lowrance P.J., McCarthy C., Becklin E.E., 1999, ApJ 512, L69
\bibitem[2000]{low1}
Lowrance P.J., Schneider G., Kirkpatrick J.D. et al., 2000, ApJ 541, L390
\bibitem[1999]{luhm99}
Luhman K., 1999, ApJ 525, 466
\bibitem[2000]{neuh00}
Neuh\"auser R., Guenther E.W., Brandner W., et al., 2000, A\&A 360, L39
\bibitem[2000]{neuh03}
Neuh\"auser R., Guenther E.W., Alves J., Hu\'elamo N., Ott Th., Eckart A., 
2003, AN 324, 535 (N03)
\bibitem[2001]{pat01}
Patience J., Ghez A.M., Reid I.N., Matthews K., 2002, AJ 123, 1570
\bibitem[2001]{reid}
Reid I.N., Burgasser A.J., Cruz K.L., Kirkpatrick J.D., Gizis J.E., 2001, AJ 121, 1720
\bibitem[2001]{reipclar}
Reipurth B., Clarke C., 2001, AJ 122, 432
\bibitem[2000]{torres2000}
Torres C.A.O., da Silva L., Quast G.R., de la Reza R., Jilinski E., 2000, AJ 120, 1410
\bibitem[2003]{torres2003}
Torres G., Guenther E.W., Marschall L.A., 2003, AJ 125, 825
\bibitem[1999]{voges}
Voges W., Aschenbach B., Boller Th., et al., 1999, A\&A 349, 389
\bibitem[1999]{webb}
Webb R.A., Zuckerman B., Platais I. et al. 1999, ApJ 512, L63
\bibitem[2001]{wuchterl01}
Wuchterl G., 2001, {\em A Dialogue on Dynamical Pre-Main Sequence Tracks}.
In: {\it The Formation of binary stars}, IAU Symp. 200, 492
\bibitem[2003]{wuchterl03}
Wuchterl G., Tscharnuter W.M., 2003, A\&A 398, 1081
\bibitem[200o]{zuckwebb}
Zuckerman B., Webb R.A., 2000, ApJ 535, 959
\end{thebibliography}
\end{document}